\documentclass[conference]{IEEEtran}
\usepackage{setspace}
\renewcommand{\baselinestretch}{0.98} \normalsize
\IEEEoverridecommandlockouts 
\usepackage{epsfig,latexsym}
\usepackage{float}
\usepackage{indentfirst}
\usepackage{amsmath}
\usepackage{amssymb}
\usepackage{times}
\usepackage{graphicx}
\usepackage{epstopdf}
\usepackage{caption}
\usepackage{subcaption}
\usepackage{cancel}
\usepackage{epstopdf}
\usepackage{ulem}
\usepackage{xcolor}
\usepackage{algorithm, algorithmic}
\usepackage{cite}
\usepackage[numbers,sort&compress]{natbib}
\begin{document}
\title{A Review of Gaps between Web 4.0 and Web 3.0 Intelligent Network Infrastructure}
\author{Zihan Zhou, Zihao Li, Xiaoshuai Zhang, Yunqing Sun and Hao Xu  
	\thanks{
		 Zihan Zhou, Zihao Li, and Xiaoshuai Zhang are with University of Glasgow, Glasgow, G12 8QQ, UK, E-mail:  \{Zihao.Li; Xiaoshuai.Zhang\}@glasgow.ac.uk; zzh45472@gmail.com. Yunqing Sun is with Department of Computer Science, McCormick School of Engineering and Applied Science, Northwestern University, Evanston, IL, US, E-mail: yunqing.sun@northwestern.edu. 
	Hao Xu is with Shanghai Engineering Research Center for Blockchain Applications And Services and Huawei Technologies, Shanghai, China, E-mail: hao.xu@ieee.org.
	}}

\maketitle 

\thispagestyle{empty}
\begin{abstract}


World Wide Web is speeding up its pace into an intelligent and decentralized ecosystem, as seen in the campaign of Web 3.0 and forthcoming Web 4.0. Marked by the Europe Commission's latest mention of Web 4.0, a race towards strategic Web 4.0 success has started. Web 4.0 is committed to bringing the next technological transition with an open, secure, trustworthy fairness and digital ecosystem for individuals and businesses in private and public sectors. 
Despite overlapping scopes and objectives of Web 3.0 and Web 4.0 from academic and industrial perspectives, there are distinct and definitive features and gaps for the next generation of WWW. In this review, a brief introduction to WWW development unravels the entangled but consistent requirement of a more vivid web experience, enhancing human-centric experience in both societal and technical aspects. Moreover, the review brings a decentralized intelligence prospect of view on native AI entities for Web 4.0, envisioning sustainable, autonomous and decentralized AI services for the entire Web 4.0 environment, powering a self-sustainable Decentralized Physical and Software Infrastructure for Computing Force Network, Semantic Network, Virtual/Mixed Reality, and Privacy-preserving content presumption. 

The review aims to reveal that Web 4.0 offers native intelligence with focused thinking on utilizing decentralized physical infrastructure, in addition to sole requirements on decentralization, bridging the gap between Web 4.0 and Web 3.0 advances with the latest future-shaping blockchain-enabled computing and network routing protocols.

\end{abstract}
\begin{IEEEkeywords} 
 Web 4.0, Web 3.0, Blockchain, Intelligence, AI, Semantic network, VR, AR, Computing Force Network.
\end{IEEEkeywords}

\newcommand{\quickwordcount}[1]{%
 
\immediate\write18{texcount -1 -sum -merge -q #1.tex output.bbl > #1-words.sum }%
 
\input{#1-words.sum} words%
}

\section{Introduction} 
Moving on from the decentralized ecosystem of Web 3.0 \cite{Xu2023}, the fourth generation of the World Wide Web has emerged through its unique requirement on intelligence and immersion between virtual and reality, known as Web 4.0, highlighted by European Commission \cite{BAHRKE2023} in the report addressing the inequality of basic rights and the interactive efficiency of the environment.
Web 4.0 is expected to combine advanced artificial and ambient intelligence, the Internet of Things, trusted blockchain transactions, virtual worlds and XR capabilities, and digital and real objects to establish an environment where every component is fully integrated and communicates with each other, enabling truly intuitive, immersive experiences, seamlessly blending the physical and digital worlds \cite{BAHRKE2023}. The EU’s strategy for Web 4.0 involves empowering users and supporting businesses in the virtual world while fostering open, inter-operable standards and multi-stakeholder governance. The ultimate goal of Web 4.0 is to pioneer user-centric, ethical and inclusive virtual worlds that boost competitiveness, foster creativity, and uphold rights.

\vspace{-0.3 em}
\subsection{Gaps between Web 4.0 and Web 3.0}

In Web 3.0, the terminology has a refined scope for decentralizing the entire World Wide Web with decentralized Applications (dApp) \cite{Xu2023,Shi2022}, decentralized physical infrastructure (DePIN) \cite{Xu2023fractal} and many blockchain infrastructures in both the network layer and the application layer \cite{Xu2023,Xu2022metaverse}.
However, Web 3.0 lacks the focus of content delivery, specifically the immersive VR/XR contents, which have exceeded the capacity of existing network infrastructure in terms of bit rates, Quality of Services (QoS) and Quality of Experience (QoE) \cite{Hewa2020,Zhang2019}. Web 4.0 sees the gap between the Web 3.0 decentralized backbone of the control plane and the incoming Web 4.0 data plane that requires network native intelligence together with future generation network infrastructures, e.g., 6G \cite{You2021}. To achieve the required data rate and connectivity of VR/XR content, semantic communication is proposed as a relief from demanding bit rates \cite{9955525,Xia2022,2019Semantic}. The Web 4.0 data is characterized as semantic data, that treat bits differently based on their features and priority \cite{Xia2022}. 

Joint Source Channel Coding (JSCC) is widely adopted in the latest research of making the semantic aware communication network \cite{2019Semantic,Xu2023semantic}. At the same time, the semantic processing leads to a computing-heavy design for future generation networks, in particular, computing force network (CFN)\cite{Li2023}, emphasizing high-performance computing with ultra-low latency and extraordinary reachability offered by both the access network \cite{Duan2021}, the core network and the data network.  


Unlike Web 3.0 which serves the same purpose in decentralized architecture, Web 4.0 introduces AI as a new entity in the network, an integrated part of the network, compared to service-only AI applications in Web 3.0. The new entity plays a pivotal role in enabling network intelligence and requires the network evolution of integrated computing and networking nodes with decentralized controllers. This model allows the AI entity to adapt, learn, and optimize itself, achieving levels of efficiency and responsiveness unattainable by the service-only AI applications in Web 3.0. Second, Web 4.0 emphasises virtual experience consumption, requiring advanced network evolution on semantic and deterministic quality of services for end consumers, as compared in Table \ref{tab:compare} with features of Web 1.0, Web 2.0. 


\begin{table*}[h]
    \centering
    \begin{tabular}{|p{1cm}|p{2cm}|p{3cm}|p{3cm}|p{3cm}|p{2cm}|p{2cm}|}
     \hline
     Web generations &Networking &Content \& Applications &Computing Infrastructure&Intelligence & Quality of Services & Security \\
     \hline
        Web 1.0 & Decentralized & Static content \& Portal page& Individual server & N/A & No QoS & N/A\\ 
        \hline
        Web 2.0 & Centralized& Interactive content \& Search engine \& Social Media & Cloud Computing \cite{Wittig2015}& N/A & Limited QoS & Certificates and PKI \cite{Adams2002}\\
        \hline 
        Web 3.0 & Decentralized \cite{Nakamoto2008} & User personalized and owned content \& Marketplace & Distributed Cloud Computing \cite{DFINITY2022}& Client-Server AI, Network for AI & Dynamic QoS & Certificateless PKI \cite{Xu2021beran}\\
        \hline
        Web 4.0 & Decentralized & Immersive media\& GPT & Computing Force Network & Semantic-aware, AI for Network & Deterministic QoE & Trust Web\\
        \hline
    \end{tabular}
    \caption{World Wide Web generations comparisons}
    \label{tab:compare}
\end{table*}

\subsection{Motivations and Contributions}
This paper contributes to Web 4.0 in three aspects: 
\begin{itemize} 
\item First, the paper gives a glance at Web history, revealing the entangled but consistent ethos of a more vivid World Wide Web with humanism, pumping enhanced virtual world interactions with considerations on privacy, efficiency and human rights. 
\item Second, a native perspective on how AI entities interact with the general public and their sustainability is detailed and envisioned. The proposed decentralized intelligence service operation principle is the key to closing the gaps in AI accessibility and enabling a self-evolving AI for Network and Network for AI by crowdsourcing and decentralized vending of AI services.  
\item Finally, the paper discusses the opportunities and challenges of regulations with an outlook on bringing a more intelligent and responsible privacy-preserving Web 4.0. 
\end{itemize}

\section{Native AI entities for Web 4}
\subsection{Decentralized operation of Computing Force Network for network native AI services}

\begin{figure}[htbp]
\vspace{-5px}
\centerline{\includegraphics[width=0.47\textwidth]{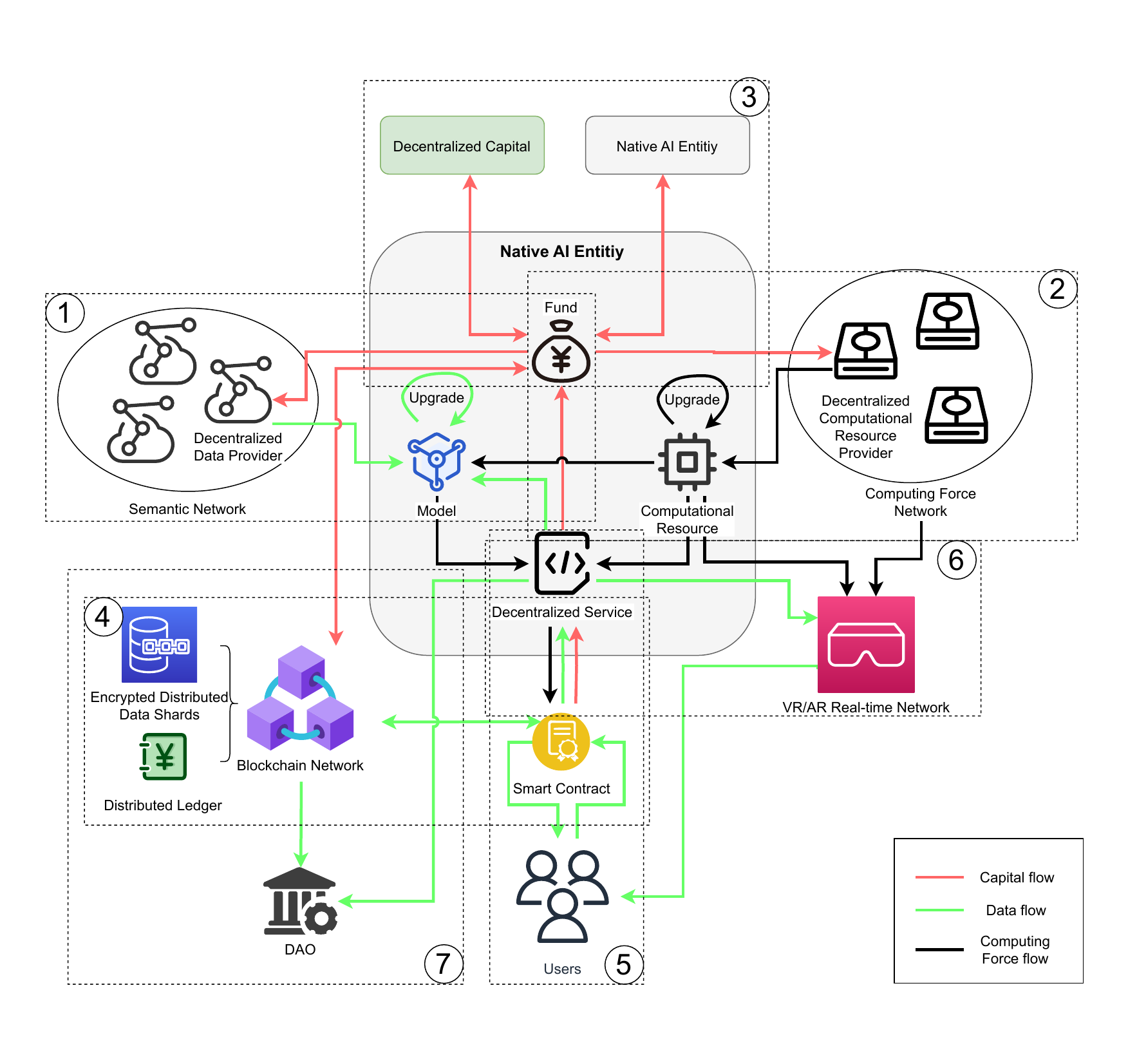}}
\caption{Framework of Native AI Entities}
\label{fig:1}
\vspace{-10px}
\end{figure}

Native AI Entities (NAEs) are autonomous artificial intelligence artifacts, purposefully designed to operate on the Web 4.0 infrastructure. Born from the collective efforts of the crowd, these NAEs are primed to serve the community that fostered their inception, with the commitment to being responsible and sustainable. Within the Web 4.0 realm, NAEs operate under a decentralized framework that is as dynamic as it is evolving. Their operational matrix intersects crucial nodes such as the Computing Force Network, Blockchain nodes, AI nodes, Semantic Networks, and the VR/AR real-time network. In Fig. \ref{fig:1}, seven main interactive scenarios and three kinds of flow are highlighted. Details of the operational blueprint of an NAE in this setting are as follows:

\paragraph{Purpose, Objectives, and Model Optimization} NAEs are designed to deliver specific AI services within the Web 4.0 environment, such as image generation or conversational interaction. They continually optimize their operations and upgrade models according to market demand, seamlessly integrating with elements like Semantic Networks and VR/AR real-time networks for an enhanced user experience. In scenario 1, NAEs use their deposited funds to compensate decentralized data providers within the Semantic Networks, obtaining the necessary training data for model optimization and upgrading.
    
\paragraph{Computational Resource} NAEs source computational power from computing service providers within the Web 4.0 network, i.e., decentralized computing services \cite{Camenisch2022,DFINITY2022} or centralized computing services, e.g., cloud computing or supercomputers. The decision to utilize specific providers is automated based on the entity's assessment of its computational requirements and service costs. NAEs have the intrinsic capability to transition or contract alternate providers when needed. As shown in scenario 2, NAEs secure the necessary computational resources from decentralized computing resource providers within the Computing Force Network, using their deposited funds as payment. This flexible approach allows NAEs to adjust their computational resources dynamically in response to changing demands.

\paragraph{Initial Funding and Financial Management}  In scenario 3, the initial funding for NAEs is crowdsourced within the Web 4.0 network, with the rules governing these funds being transparently communicated to all contributors. NAEs can autonomously manage their budgets and adjust their financial strategies based on real-time analysis and anticipated future needs. The deposited funds of an NAE may circulate to other NAEs or decentralized capital within the Web 4.0 network, forming a complex and dynamic network of capital circulation. This financial ecosystem is adaptive and responsive, promoting financial flexibility and resilience, and accommodating the unique requirements of various NAEs while facilitating collaboration and interdependence among them. NAEs prioritize efficiency, transparency, and accountability in their financial management, ensuring that all financial operations support the delivery of their AI services and contribute to the overall sustainability of their operations.

\paragraph{Blockchain Network Interaction} NAEs interact with the blockchain network primarily via smart contracts in scenario 4. This process represents a major aspect of the financial and data management activities within the Web 4.0 network. Deposited funds from NAEs may flow into the blockchain network, which in turn, can channel funds back to the NAEs. The implementation of blockchain offers a secure and transparent method for managing finances and data, fostering trust and credibility in the operations of NAEs while enabling real-time tracking and verification of transactions.

\paragraph{Provision of Decentralized Services and User Data Protection} In scenario 5, users can call NAEs' decentralized services via smart contracts. This process involves the flow of funds and data from the user to the NAEs and the provision of computational resources and data to the user. NAEs operate in compliance with data protection regulations, requiring explicit user permissions to utilize user data. The process of securing permissions and usage of the data is recorded and verifiable on the blockchain, ensuring privacy and security.

\paragraph{VR/AR Real-time Network Interaction} In the sixth interaction scenario, the computational resources of an NAE and the Computing Force Network collectively provide computational power to the VR/AR Real-time Network, as detailed in \cite{Xu2023semantic}. NAEs also supply data obtained from their decentralized services to the VR/AR network, enabling real-time virtual or augmented reality experiences.

\paragraph{Risk Assessment and DAO Audit} Investors are made aware of the inherent risk in NAE operations. To facilitate risk assessment, details of NAE operations, including service quality and financial status, are made transparent on the blockchain network. In the seventh interaction scenario, decentralized autonomous organizations (DAOs) audit NAEs' services and capital flow, ensuring service quality, fairness, and accountability.

\subsubsection{Four Phases: Initialization, Growth, Steady and Retirement}

\begin{figure}[htbp]
\vspace{-5px}
\centerline{\includegraphics[width=0.47\textwidth]{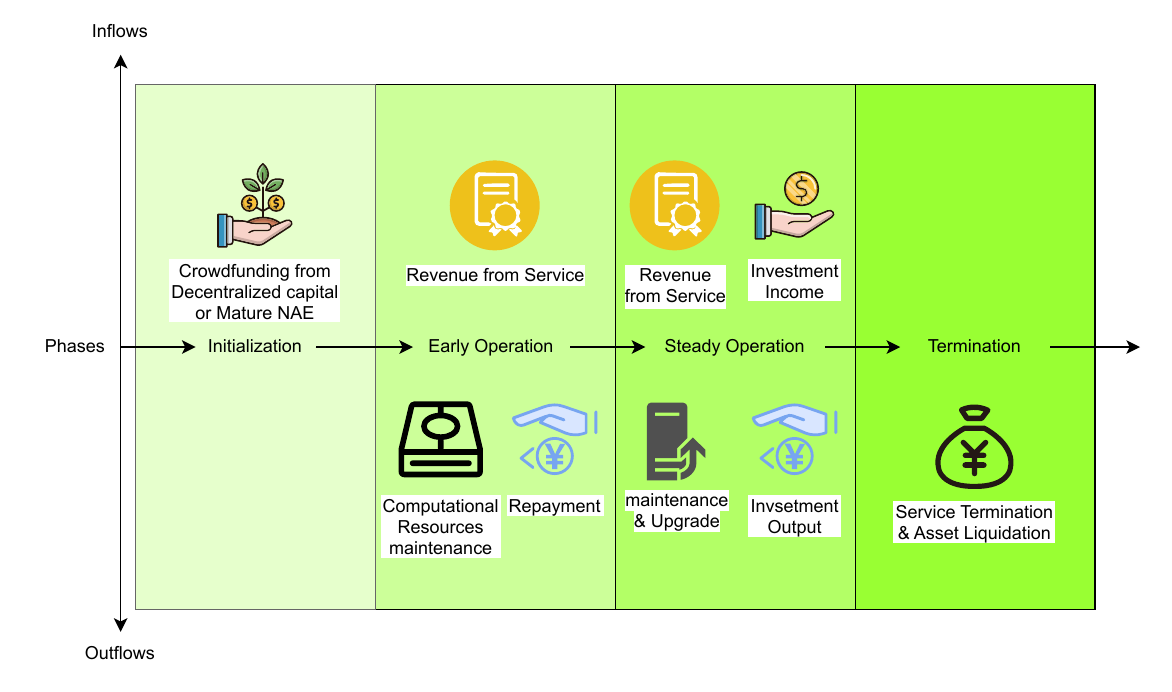}}
\caption{Operation phases of Native AI Entities}
\label{fig:2}
\vspace{-10px}
\end{figure}

As autonomous constructs within the Web 4.0 environment, NAEs operate under a life-cycle model to facilitate specific AI services. This life cycle encompasses four stages: Initialization and Configuration, Early Operation and Growth, Steady Operation and Expansion, and Decline and Termination. The progression through these stages is marked by distinctive operational behaviors and economic transactions, revealing how an NAE acquires, utilizes, and manages its resources in pursuance of its service objectives. Each phase also presents a snapshot of the economic flows at various life-cycle stages, shown in Fig. \ref{fig:2}, from initial capital accumulation to resource allocation, revenue generation, reinvestment, and eventually, asset liquidation.

\begin{itemize}
\item Phase 1: Initialization and Configuration:
\begin{enumerate}
    \item Acquisition of Capital and Computational Resources: NAEs generate their initial funds via crowdfunding efforts or from mature NAEs, employing these funds to lease computational resources.
    \item Market Positioning: This involves identifying service provisions, target markets, and potential users.
    \item Resource and Market Analysis: An efficiency assessment of resource allocation is performed along with predictions for market demands and potential gains.
\end{enumerate} 
\item Phase 2: Early Operation and Growth:
\begin{enumerate}
    \item Provision of Services and Revenue Generation: NAEs commence their operations by providing services. Revenue is generated via these services and the execution of smart contracts.
    \item Repayment to Initial Investors: A portion of the revenue is deducted by the NAE to repay the initial investors.
    \item Market Expansion: Market feedback is analyzed and service provision adjusted while exploring new service markets and growth points.
\end{enumerate} 
\item Phase 3: Steady Operation and Expansion:
\begin{enumerate}
    \item Stable Revenue and Self-Sustenance: By providing services and generating revenue, NAEs attain self-sustenance and gradually repay the initial investors.
    \item Service Optimization and Innovation: Existing services are improved and optimized, which may include new AI models, superior user experiences, etc.
    \item Investment in Other NAEs: Surplus revenue is used to invest in other budding NAEs or blockchain projects for additional profit.
    \item Economic Benefit Analysis: Economic benefits are analyzed, optimizing investment and operation strategies.
\end{enumerate} 
\item Phase 4: Sunset and Termination:
\begin{enumerate}
    \item Service Termination and Asset Liquidation: NAEs cease their services and liquidate any remaining assets.
    \item Investor Refunds: Any remaining assets (if any) are returned to the investors.
    \item Lessons and Experience: The causes of failure are analyzed, offering valuable lessons and experiences for other NAEs and investors.
\end{enumerate} 
\end{itemize}

\subsection{Semantic content aware network for VR/AR in confidential peer-to-peer network}

Semantic Networks play an essential role within NAEs, allowing them to understand and process complex information across diverse contexts. By mapping relationships and drawing inferences between different entities, NAEs can generate context-relevant responses and provide highly personalized services in the VR/AR space. For instance, an NAE serving as a VR guide could understand a user's interests and pre-train JSCC models based on the characteristics of content \cite{Xu2023semantic}.

As NAEs process complex and often sensitive information, they employ robust measures to prevent potential misuse of this information within the network. A key component of these preventative measures is confidential computing, which ensures that data remains encrypted while being processed. This secure processing prevents unauthorized access, safeguarding the information even when operating in shared or potentially insecure environments.

In addition to confidential computing, NAEs are considered operated with the support of Trusted Execution Environments (TEE) to further enhance data security. TEE creates isolated, secure environments for processing sensitive information away from the rest of the system \cite{Xu2022metaverse}. Even in scenarios where a malicious entity gains access to the network, the integrity and confidentiality of the data being processed within the TEE remain unaffected.
Beyond these protective measures, NAEs also leverage their AI capabilities for real-time network monitoring and anomaly detection. This allows them to swiftly identify potential threats and take immediate corrective actions against abnormal traffics.

\section{Opportunities and Challenges}

\subsection{Privacy challenges in identity and data distribution} 
Since Web 4.0 is evolved from Web 3.0, it inherits the privacy challenges from not only Web 3.0 itself but also the shift from Web 1.0/2.0 to Web 3.0. To transform the centralized identity management in Web 1.0/2.0 to a decentralized manner, a universal identity management strategy is proposed in Web 3.0 using zero-knowledge proof (ZKP), multi-party computation (MPC) and other cryptographic methods \cite{Xu2023, Xu2023fractal}. Such a strategy utilizes ZKP to protect users' real identities and derive different identities used by networks and applications in Web 3.0. Furthermore, the derived identities can be publicly verified by smart contracts but the private information related to the identities is not revealed. However, this rudimentary design for Web 3.0 does not consider the border control issue in identity verification. Since its nationalities issue the real identities of a user, the derived identities based on the real identities can be only verified by the networks and applications in the countries of the user's nationalities with the assistance of the corresponding authorities. When a user needs to access networks and applications outside its nationalities, the identity verification performed by other countries still needs to be further advanced in Web 4.0. Specifically, the interaction between two or more smart contracts of identity verification belonging to different countries should be considered and designed. In addition, the private information of users involved in such abroad identity verification also needs to be standardized to protect user privacy and ease the identity verification process to ensure desirable access efficiency in Web 4.0.

``Read-write-own'' is the core of Web 3.0, representing the self-governance of data for all users in equivalent. In Web 3.0, all users aim to freely reach content, services and applications deployed in decentralized networks but not in centralized servers applied in Web 1.0/2.0. Meanwhile, all users can manage their data in distributed storage and authorize access without barriers. As the next generation of the Internet inherits and further develops Web 3.0, Web 4.0 should keep building user data autonomy. However, more and more NAEs involved in Web 4.0 networks, content, services and applications may challenge the progress of data autonomy. Compared with Web 3.0, a distinctive characteristic of Web 4.0 is tremendous data generated by NAE for different purposes. Although there is a multi-layer identity strategy has been suggested to enable users to use different identities in different scenarios in Web 3.0, fine-grained permissions and multi-layer tags for different parts of data have not been considered for users to control who can access specific data, especially for massive data generated by AI. In addition, data flows and content distributions across different countries may incur censorship difficulty for authorities due to varied data censorship mechanisms and standards. On the other hand, censorship conflicts with data autonomy in users' view since censorship may leak user private information. Therefore, data censorship by different countries should be regulated in a uniform manner with restricting the exposure of sensitive parts of data.

\subsection{Regulatory opportunities and concerns} 
As discussed in previous sections, NAE-enabled Web 4.0 can benefit individual by achieving legal objectives in cyberspace. The existing legislative framework in cyberspace already applies to several aspects of Web 4.0; however, there is no a powerful enforcement toolkit. In relation to users' privacy and personal data, General Data Protection Regulation (GDPR) establishes an omnibus system of informed-consent-based obligations and rights \cite{GDPR}. 


The advent of NAE and the transition to Web 4.0 have fundamentally revolutionized the ways users interact with service providers. For example, with NAE's assistance, users now have an enhanced capacity to review privacy policies in a more comprehensive manner. Users can utilize the automated NAE mapping to gain a clear understanding of the types of personal data the enterprises are collecting, processing, and storing. Meanwhile, NAEs can be tailored to learn the users' privacy preferences in order to better protect their privacy. In this way, the service can be personalized without sacrificing users' privacy in different contexts. This technology not only enables individuals to exercise their data subject rights, such as access, erasure, and rectification in a timely manner, but it also aids in detecting and flagging potential data breaches. Consequently, NAE-enabled Web 4.0 helps meet notification requirements, offering users a level of data security and privacy protection that was previously unattainable \cite{Yu2021shard}.

Moreover, NAE-enabled Web 4.0 can provide valuable analytical support for conducting data protection impact assessments required under data privacy regulations like the GDPR. It can systematically categorize and analyze personal data flows, flag potential compliance issues around legal basis or data minimization, identify inherent risk factors such as large-scale profiling or automated decision systems, assess potential impacts on individuals, and suggest technical controls to mitigate risks. 

As for the individuals' protection and regulations on competition and innovation, the Digital Services Act (DSA) and the Digital Markets Act (DMA) also apply to Web 4.0. In this context, NAE could automatically monitor online service providers for the presence of illegal content or prohibited goods. This could enable timely flagging and removal as well as enforcement of notice-and-action procedures. 
Regarding the anti-competitive practices, NAE-enabled Web 4.0 could identify anomaly anti-competitive activities that contravene DMA rules in cyberspace. For instance, self-preferencing and exclusion of competitor products or services can be automatically flagged to regulators. At a technical level, NAE testing and analysis of dominant service providers' APIs, data structures, and interoperability presents the means to evaluate and ensure compliance with data portability obligations under the DMA. Such NAE functionality applied in continuous compliance processes promotes the DMA's goals in regulating the gatekeepers on digital ecosystems. However, oversight and evaluation of NAE's reasoning is critical to ensure appropriate recommendations aligned with regulatory goals.


While promising, more advanced legal and ethical oversight is required to align NAE in Web 4.0 with our fundamental values and rights protection in whole decentralized ecosystems. Existing conflicts between this emerging web architecture and legislations can been foreseen. For example, there are some potential conflicts between Right to be Forgotten (RtbF), granted by the GDPR and many other data privacy legislations, and the NAE-enable system \cite{Li2021social}. A core tension arises from the fact that personal data used to train AI systems can become deeply embedded within the models' architectural parameters, neural network weights, and decision-making logic flows. Unlike data residing in traditional databases, deleting or removing the original training data does not necessarily erase its latent traces from the AI model itself. In a sense, the model persists as a record of its own training. This presents challenges in fully implementing individuals' right to be forgotten requests and erasing their personal data under GDPR Article 17 when such data was utilized in model development.

Confronting these concerns and rapidly evolving technologies, regulatory experimental tools are a valuable approach to collecting data, assessing legal, institutional and technological ramifications and shaping new regulatory approaches outside of prevailing regulatory frameworks. Such experimental regulatory approaches include regulatory sandboxes, standardization, and co-regulation involving regulators, industrial guidelines and markets. Among these, sandboxes create a controlled area where participators obtain a waiver from certain legal provisions and compliance processes and attain tailored legal support to foster the development of new technologies. There are several successful instances for setting up such experimental regulatory environments. The UK Financial Conduct Authority (FCA) pioneered the first fintech regulatory sandbox \cite{FCA2015}, while the UK Information Commissioner’s Office is testing the impact of AI-related products and services on data privacy frameworks \cite{InformationCommissionersOffice2023}. 

Therefore, similar experimental regulatory toolkit can be used in the context of Web 4, in order to provide appropriate flexibility while protecting legal principles and fundamental rights. However, such experimental regulation requires careful design and testing. Otherwise, a lack of harmonized and standardized criteria, testing processes and inadequate specifications will harm competition, consumers, and public or personal data. In the context of Web 4, crafting the experimental regulatory environment firstly necessitates a multidisciplinary regulatory framework, given the convergence of various emerging technologies inherent in Web 4.0. 
Secondly, multi-stakeholder’s interests need to be considered. As Web 4.0 encompasses a complex ecosystem and impacts various sectors, the diverse interests of all stakeholders should be taken into account, including enterprises, SMEs, authorities, individuals, operators, standardization bodies and legislators, among others. These necessitate a strong regulatory interoperability in an international level.
Thirdly, the comprehensive eligibility and testing criteria for experimental regulation are crucial. On the one hand, the pro-innovation mechanisms are important in order to provide flexibility. On the other hand, the design of testing parameters, duration, entry requirements and terminal conditions are also essential. Therefore, in the context of Web 4, the experimental regulatory tools should be carefully designed to embody the aforementioned concerns and probe into the potential impact on the risks on users’ protection and innovation.

\section{Conclusions and Future Work}
In this paper we summarise the key impacts of Web 4.0 and detail how it is different from Web 3.0 in a tight scope developed within Europe Commissioners' outlook emphasizing not only the intelligence, decentralization, semantic and VR/XR ready but also regulator improvements on human rights, privacy and sustainability. A novel NAE operation model is proposed to benefit the network with its native ecosystem of sustainable AI and provides ubiquitous AI to every aspect of network, e.g., Autonomous Driving Network, NetGPT, ChatGPT-alike service, etc. Meanwhile, the end-to-end semantic communication network is also projected to be the key enabler for Web 4.0 due to the demands of VR/XR ready services with assured Quality of Services. In the end, this paper provides insight into regulator opportunities regarding the privacy, content ownership, sustainability, legal and ethics perspectives of Web 4.0.

\normalem
\bibliographystyle{IEEEtran}
\bibliography{bcmag}

\end{document}